\documentclass[aps,prl,reprint,graphicx,groupedaddress]{revtex4-1}

\usepackage{graphicx}
\usepackage{amssymb}

\begin{document}



\title{An iterative algorithm to improve colloidal particle locating}

\author{K. E. Jensen} 
\email[]{kjensen@post.harvard.edu}
\affiliation{Department of Physics, Harvard University, Cambridge, Massachusetts, 02138, USA}
\affiliation{Department of Mechanical Engineering and Materials Science, Yale University, New Haven, CT, 06511, USA}

\author{N. Nakamura} 
\affiliation{Graduate School of Engineering Science, Osaka University, Toyonaka, Osaka 560-8531, Japan}

\date{\today}

\begin{abstract} 
Confocal microscopy of colloids combined with digital image processing has become a powerful tool in soft matter physics and materials science. 
Together, these techniques enable locating and tracking of more than half a million individual colloidal particles at once. 
However, despite improvements in locating algorithms that improve position accuracy, it remains challenging to locate \textit{all} particles in a densely-packed, three dimensional colloid without erroneously identifying the same particle more than once. 
We present a simple, iterative algorithm that mitigates both the ``missed particle'' and ``double counting'' problems while simultaneously reducing sensitivity to the specific choice of input parameters. 
It is also useful for analyzing images with spatially-varying brightness in which a single set of input parameters is not appropriate for all particles. 
The algorithm is easy to implement and compatible with existing particle locating software.
\end{abstract}

\maketitle

Colloids are materials comprised of nanometer- to micrometer-scale solid particles suspended in a fluid.
Colloidal particles can easily be made to assemble into a variety of structures and phases, including gels, crystals, and glasses \cite{Davis1987, Poon1997, Lu2013}, which can then be manipulated locally or as macroscopic materials. 
Three dimensional confocal microscopy of colloids combined with fast computers and precise image processing makes it possible to track individual particle locations over time in a bulk colloid \cite{Dinsmore2001, Prasad2007}. 
These techniques enable precise measurement of the structure and dynamics of materials over the entire range of length scales from the constituent particles up to the bulk. 
They have been used for investigations into colloid structure formation \cite{vanBlaaderen1997, Gasser2001, Schofield2005, Gasser2009, JensenSM2013}, gel structure and dynamics \cite{Dinsmore2002, Lu2008, Sprakel2011, KodgerThesis2015}, dynamical heterogeneities and local deformation mechanisms in glasses \cite{Dinsmore2001, Poon2007, Schall2006, Jensen2014}, aging of glasses \cite{glass_aging}, and defects and elasticity in crystals \cite{Schall2004, Schall2007, MariaPGThesis2012, JensenSM2013, Russell2015}. 

However, the information to be gained from these experiments is only as accurate as the particle locations themselves, since they are the foundation upon which all further analyses are based. 
For this reason, it is essential that the particles be located as precisely, accurately, and completely as possible. 
There exist several general approaches to object detection in images \cite{Ratches2011, Kerekes2008}, but these methods are usually not accurate enough for colloidal particle location \cite{Irvine2010}.
Standard algorithms do exist for the specific problem of precisely locating spherical objects in 3D images, and which address the particular problem of finding particles in colloidal suspensions \cite{CrockerGrier}. 
Several recent studies have focused on algorithms to improve the accuracy and precision of particle locating \cite{Lu2007, Jenkins2008, Gao2009, Lu2013_2}. 
However, significant challenges remain, particularly in analyzing 3D images of densely-packed particles or images with spatially-varying brightness.
While individual particles may be located very accurately, it is very common for some particles either to be missed entirely or identified more than once (``double-counted''). 
A user can avoid double-counted particles either by enforcing a strict minimum separation between particle locations or by adjusting input parameters to lower the sensitivity of the locating software, but at the cost of missing more particles.
On the other hand, increasing the software sensitivity also increases the number of particles found multiple times;
although setting a separation threshold works fairly well to limit this double-counting, it introduces an additional parameter that may strongly affect final results and there is still no guarantee that all particles will be found. 
The result is a tradeoff between these two types of errors that is very sensitive to the user's precise choice of input parameters. 
Attempts to choose input parameters that mitigate both problems simultaneously inevitably produce some of each type of error.

To solve this problem, we developed an iterative particle locating algorithm that is able to find all of the particles in a 3D image of a colloid with few or no double-counted particles. 
We first run standard particle locating software on the original image data, using input parameters that deliberately err on the side of missing particles in order to avoid double-counted particles.
Next, we erase from the original image those particles that have already been found.
This creates a new ``raw'' residual image that is empty except for those particles that have not yet been found.
We then apply the particle location software to the residual raw image, and iterate this procedure until all particles have been found. 

\begin{figure}[h]
\includegraphics[ bb= 0 0 246 410]{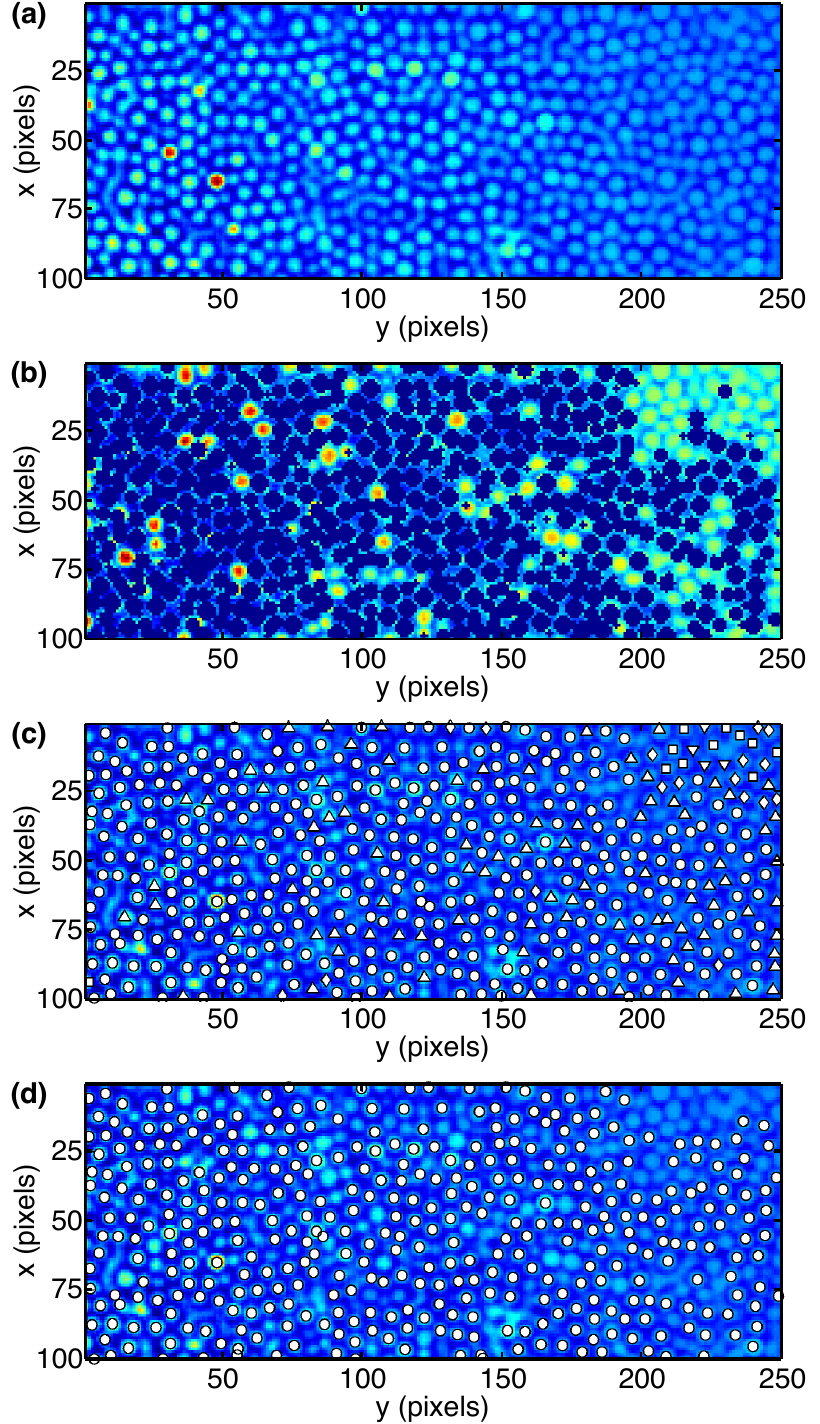}
\caption{\label{iterative_locating_figure} Iterative particle locating applied to a 3D confocal image of a dense, disordered colloid with variation in individual particle brightness and uneven illumination across the sample \cite{supplement}. 
(a) An x-y cross section through the raw data. 
(b) The same cross section through the residual raw data after particles found in the first locating pass have been deleted. 
(c) The results of iterative particle locating. Particles whose centers are within about $\pm 1$ particle radius of this cross section are marked according to the locating iteration in which they were found: 1st pass, ($\circ$); 2nd pass, ($\bigtriangleup$); 3rd pass, ($\diamond$); 4th pass, ($\square$); 5th pass, ($\bigtriangledown$). 
(d) The results for the same 3D data set of standard non-iterative particle locating with optimized parameters.}
\end{figure}

The steps of the iterative algorithm are described in detail below, illustrated for an example image stack in Figure \ref{iterative_locating_figure}. 
For a typical 3D image containing 50,000 particles of equal brightness, we locate about 95\% of the particles on the initial pass through the data, and the iterative locating is complete after four to six iterations. 
At the end, we estimate that $<$0.2\% of the particles are double-counted, and these isolated pairs can easily be identified and consolidated.
For a given sample, algorithm performance will depend on the resolution and quality of the original image data; 
for example, images with spatially-varying brightness, such as the example shown in Figure \ref{iterative_locating_figure}, often require more iterations than evenly-illuminated samples.

The software we use to implement iterative particle locating is included as MATLAB code in the Supplement \cite{supplement}.
We also provide three example data sets: Example 1, a colloidal glass; Example 2, a colloidal crystal; and Example 3, a dense, disordered colloid with spatially-varying image brightness (used for the example shown in Figure \ref{iterative_locating_figure}).
Each example data set includes a small 3D confocal image stack, a text file of particle locations, and the input parameters used to locate the particles. 
We interface the iterative algorithm with the publicly-available MATLAB particle locating software of Ref. \cite{Gao2009}. 
However, the algorithm we present is broadly compatible with any approach to particle locating, and the example code provided should be easily adaptable to interface with other locating software.

We start with a 3D confocal image of a colloid in which the particles appear bright against a dark background.
An x-y cross section through such an image is shown in Figure \ref{iterative_locating_figure}(a).
We bandpass filter the raw image and run a first pass of particle locating using the standard filtering and feature-finding software of Ref. \cite{Gao2009}. 
In choosing input parameters for the standard software, we deliberately choose parameters that may miss some particles rather than double-counting any particle. 
Any missed particles will be identified in subsequent iterations.

Next, we generate a 3D residual raw image by deleting from the original image those particles that have already been found.
Spherical particles generally appear as ellipsoids in 3D images, so we erase each found particle by replacing the original image data surrounding its coordinates with an appropriately-sized ellipsoid of zero-valued pixels.
The x, y, and z particle dimensions in pixels are the only input parameters required by the iterative locating algorithm beyond standard particle locating.

The resulting residual raw image contains only those particles that have not yet been found, as shown for the example cross section in Figure \ref{iterative_locating_figure}(b). 
In this case, a number of particles were missed during the first pass.
Particles that were missed are usually isolated in the residual raw image, as on the left side of the example, making them much easier to locate on subsequent iterations than when they were surrounded by other bright objects.
In cases where spatially-varying image brightness causes the initial locating pass to miss entire regions of particles, as in the upper-right corner of the example, the iterative locating procedure will locate all of the particles over the course of several iterations.

We then bandpass filter the residual raw image and run the particle locating software again.
We use all the same input parameters for filtering and locating as in the first pass, with the added constraint that a bright region must have an integrated intensity greater than some minimum threshold in order to be considered a real particle rather than noise.
This is a standard feature of many particle locating software packages, including the one we use \cite{Gao2009}.
This cutoff prevents any small bright regions in the residual raw image from being falsely identified as additional particles, such as the edges of particles that may have been incompletely deleted from the original raw image. 
As these regions are significantly smaller than the real particles, there is no ambiguity in distinguishing them.
Particles cut off by the edges of the image may or may not pass the integrated intensity threshold.
Although they are real particles, their center coordinates will not be accurate if they are not fully contained within the image.
Fortunately, it is straightforward to exclude particles close to the image edges from subsequent analyses.

We continue to delete particles from the images as they are located, and iterate this entire process until no new particles are found.
For a typical image with uniform brightness, the second iteration finds nearly all of the particles missed during the first pass, and the locating is usually complete in 4-6 iterations.
Samples that were unevenly illuminated by the microscope or that have a large variation in the brightness of individual particles may require more iterations.
These are also data sets for which a single set of locating parameters would usually not successfully identify all particles, so the iterative approach is particularly helpful here.
The 3D data set shown in the Figure required 8 iterations to complete the locating.

Figure \ref{iterative_locating_figure}(c) shows all of the particles located near this cross-section at the end of the iterative locating process.
Each found particle is marked according to the iteration in which it was located.
For comparison, we also show the results of traditional single-pass particle locating in Figure \ref{iterative_locating_figure}(d).
In this case, we optimized the parameters to locate as many particles as possible in a single pass with minimal double-counting, including setting a minimum separation distance between neighboring particles.
However, the locating results are only marginally more complete than the first pass of the iterative particle locating, and many particles in this example were missed entirely.

In summary, we have created and implemented an iterative algorithm to improve the completeness particle locating of individual colloidal particles in an image. 
The algorithm is useful, simple, straightforward to implement, and easily integrated with existing particle locating software. 
The algorithm requires only three new input parameters from the user: the x, y, and z dimensions in pixels of an individual particle in the original image.
Although we implement it for locating monodisperse, spherical particles, this algorithm could easily be extended to colloids comprised of polydisperse or non-spherical particles.
In the latter case, the orientation of the already-found particles would also be required to erase them accurately from the raw images.

This work was supported by NSF through the Harvard MRSEC (Contract \#DMR-1420570). We thank Daniel Pennachio for help imaging the colloidal crystal included as Example 2 in the Supplement, and Peter Schall, Sanne van Loenen, and Triet Dang for providing the image data for Example 3 in the Supplement and the Figure.  We are also grateful to John Irvine and Frans Spaepen for helpful discussions, and to the Reviewer for useful suggestions that improved the manuscript.


\begin{thebibliography}{15}

\bibitem{Poon1997} W. C. K. Poon and M. D. Haw, Adv. Colloid Interface Sci. \textbf{73}, 71 (1997).

\bibitem{Davis1987} K. E. Davis and W. B. Russel, Adv. Ceramics \textbf{21}, 573 (1987).

\bibitem{Lu2013} P. J. Lu and D. A. Weitz, Annu. Rev. Cond. Matter Phys. \textbf{4}, 217 (2013).

\bibitem{Dinsmore2001} A. D. Dinsmore, E. R. Weeks, V. Prasad, A. C. Levitt, and D. A. Weitz, Appl. Opt. \textbf{40}, 4152 (2001).

\bibitem{Prasad2007} V. Prasad, D. Semwogerere,  and E. R. Weeks, J. Phys. Cond. Matter \textbf{19}, 113102 (2007).

\bibitem{vanBlaaderen1997}A. van Blaaderen, R. Ruel,  and P. Wiltzius, Nature \textbf{385}, 321 (1997).

\bibitem{Gasser2001} U. Gasser, E. R. Weeks, A. Schofield, P. N. Pusey, and D. A. Weitz, Science \textbf{292}, 258 (2001).

\bibitem{Schofield2005} A. B. Schofield, P. N. Pusey, and P. Radcliffe, Phys. Rev. E \textbf{72}, 031407 (2005).

\bibitem{Gasser2009} U. Gasser, J. Phys. Cond. Matter \textbf{21}, 203101 (2009).

\bibitem{JensenSM2013} K. E. Jensen, D. Pennachio, D. Recht, D. A. Weitz, and F. Spaepen, Soft Matter \textbf{9}, 320 (2013).

\bibitem{Dinsmore2002} A. D. Dinsmore and D. A. Weitz, J. Phys. Cond. Matter \textbf{14}, 7581 (2002).

\bibitem{Lu2008} P. J. Lu, E. Zaccarelli, F. Ciulla, A. B. Schofield, F. Sciortino, and D. A. Weitz, Nature \textbf{453}, 499 (2008).

\bibitem{Sprakel2011} J. Sprakel, S. B. Lindstr\"om, T. E. Kodger, and D. A. Weitz, Phys. Rev. Lett. \textbf{106}, 248303 (2011).

\bibitem{KodgerThesis2015} T. E. Kodger, Mechanical Failure in Colloidal Gels, Ph.D. Thesis, Harvard University, 2015.

\bibitem{Poon2007} R. Besseling, E. R. Weeks, A. B. Schofield,  and W. C. K. Poon, Phys. Rev. Lett. \textbf{99}, 028301 (2007).

\bibitem{Schall2006} P. Schall, I. Cohen, D. A. Weitz, and F. Spaepen, Nature \textbf{440}, 319 (2006).

\bibitem{Jensen2014} K.E. Jensen, D.A. Weitz, and F. Spaepen, Phys. Rev. E \textbf{90}, 042305 (2014).

\bibitem{glass_aging} G. C. Cianci, R. E. Courtland, and E. R. Weeks, Solid State Commun. \textbf{139}, 599 (2006).

\bibitem{Schall2004} P. Schall, I. Cohen, D. A. Weitz, and F. Spaepen, Science \textbf{305}, 1944 (2004).

\bibitem{Schall2007} P. S. Schall, D. A. Weitz, and F. Spaepen, Science \textbf{318}, 1895 (2007).

\bibitem{MariaPGThesis2012} M. C. M. Persson Gulda, Defects in hard-sphere colloidal crystals, Ph.D. Thesis, Harvard University, 2013.

\bibitem{Russell2015} E. R. Russell, F. Spaepen, and D. A. Weitz, Phys. Rev. E \textbf{91}, 032310 (2015).

\bibitem{Ratches2011} J. A. Ratches, Opt. Eng. \textbf{50}, 072001 (2011).

\bibitem{Kerekes2008} R. A. Kerekes and B. V. K. V. Kumar, Opt. Eng. \textbf{47}, 067202 (2008).

\bibitem{Irvine2010} J. M. Irvine, GeoTech 2010, Sponsored by the American Society of Photogrammetry and Remote Sensing (ASPRS), 27-28 September 2010, Fairfax, VA.

\bibitem{CrockerGrier} J. C. Crocker and D. G. Grier, J. Colloid Interface Sci. \textbf{179}, 298 (1996). Particle locating software based on this reference is available at http://www.physics.emory.edu/faculty/weeks/idl/

\bibitem{Lu2007} P. J. Lu, P. A. Sims, H. Oki, J. B. Macarthur, and D. A. Weitz, Opt. Express \textbf{15}, 8702 (2007).

\bibitem{Jenkins2008} M. C. Jenkins and S. U. Egelhaaf. Adv. Colloid Interface Sci. \textbf{136}, 65 (2008).

\bibitem{Gao2009} Y. Gao and M.L. Kilfoil, Opt. Express \textbf{17}, 4685 (2009). Software available under ``MATLAB 3D feature finding algorithms'' at http://people.umass.edu/kilfoil/downloads.html

\bibitem{Lu2013_2} P. J. Lu, M. Shutman, E. Sloutskin, and A. V. Butenko, Opt. Express, \textbf{21}, 30755 (2013).


\bibitem{supplement} See supplemental material for an example MATLAB implementation of the iterative algorithm as well as three small example data sets, each with a text file of particle positions and a .mat file of the input parameters used to locate the particles. Supplemental material is currently available at http://campuspress.yale.edu/kej4/iterative/





\end{thebibliography}
\end{document}